\documentclass[runningheads]{llncs}
\usepackage{color}
\usepackage{graphicx}
\begin{document}
\title{AutoPET Challenge 2022: 
Step-by-Step Lesion Segmentation in Whole-body FDG-PET/CT}
\titlerunning{AutoPET Challenge 2022}
%
\author{Zhantao Liu\orcidID{0000-0002-3934-2660} \and 
Shaonan Zhong\orcidID{0000-0003-1648-1489} \and 
Junyang Mo\orcidID{0000-0002-2981-9349}}
\authorrunning{Liu et al.}
%
\institute{School of Biomedical Engineering, 
Shenzhen University, Shenzhen Guangdong 518055, China}
\maketitle            
\begin{abstract}
Automatic segmentation of tumor lesions is a critical initial processing step for quantitative PET/CT analysis. However, numerous tumor lesions with different shapes, sizes, and uptake intensity may be distributed in different anatomical contexts throughout the body, and there is also significant uptake in healthy organs. Therefore, building a systemic PET/CT tumor lesion segmentation model is a challenging task. In this paper, we propose a novel step-by-step 3D segmentation method to address this problem. We achieved Dice score of 0.92, false positive volume of 0.89  and false negative volume of 0.53  on preliminary test set.The code of our work is available on the following link: https://github.com/rightl/autopet. 

\end{abstract}
\section{Introduction}
Positron Emission Tomography/Computed Tomography (PET/CT) is an integral part of the diagnostic workup for various malignant solid tumor entities. Due to its wide applicability, Fluorodeoxyglucose (FDG) is the most widely used PET tracer in an oncological setting reflecting glucose consumption of tissues, e.g. typically increased glucose consumption of tumor lesions.

As part of the clinical routine analysis, PET/CT is mostly analyzed in a qualitative way by experienced medical imaging experts. Additional quantitative evaluation of PET information would potentially allow for more precise and individualized diagnostic decisions.

A crucial initial processing step for quantitative PET/CT analysis is segmentation of tumor lesions enabling accurate feature extraction, tumor characterization, oncologic staging and image-based therapy response assessment. Manual lesion segmentation is however associated with enormous effort and cost and is thus infeasible in clinical routine. Automation of this task is thus necessary for widespread clinical implementation of comprehensive PET image analysis.

Recent progress in automated PET/CT lesion segmentation using deep learning methods has demonstrated the principle feasibility of this task. However, despite these recent advances tumor lesion detection and segmentation in whole-body PET/CT is still a challenging task. The specific difficulty of lesion segmentation in FDG-PET lies in the fact that not only tumor lesions but also healthy organs (e.g. the brain) can have significant FDG uptake; avoiding false positive segmentations can thus be difficult.

In this study, we propose a 3D U-Net step-by-step segmentation method to address the challenge of whole-body tumor lesion segmentation.

\section{Method}

\subsection{Model structure}
Our model is a standard U-net \cite{ref_article1} implemented by MONAI. The input channels are PET and CT images, and the output has eight channels. The depth of the U-Net is 5. The number of channels in each block is 96, 192, 384, 768, and 1536. The number of residual units is 4.
\subsection{Step-by-Step Segmentation} 

\begin{figure}
\includegraphics[width=\textwidth]{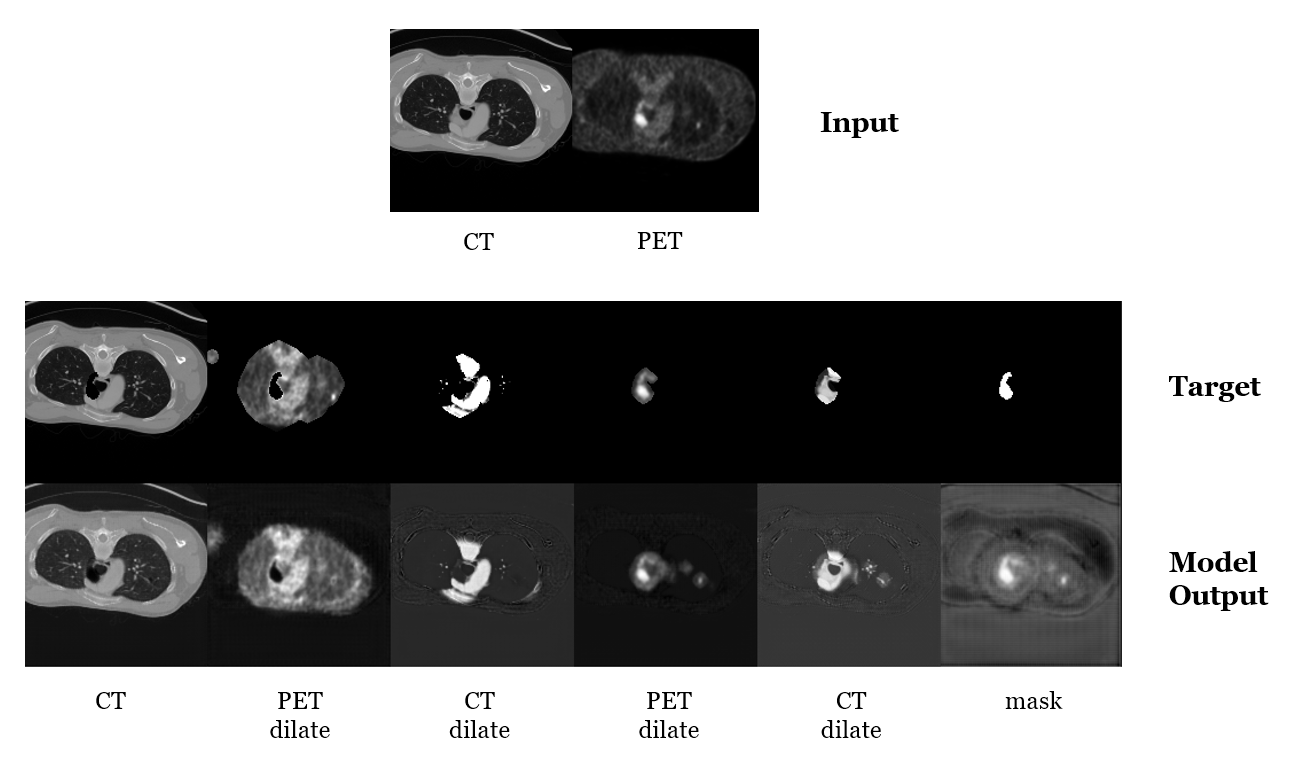}
\caption{U-Net input and output channels of our model. For the output channels, we dilate the segmentation mask several times and fill the masked area with input images.} \label{fig1}
\end{figure}

In Kojima, Takeshi, et al's \cite{ref_article2} work, they use prompts to guide large language models to answer a question step-by-step. In this way, the zero-shot accuracy of models like GPT-3 is improved by more than 70 percent. Inspired by their work, we apply the idea of step-by-step thinking to our segmentation task.

As shown in \textcolor{red}{Fig. 1}, our U-Net outputs eight channels, one of which is a regular segmentation mask and the others are auxiliary channels. To achieve step-by-step segmentation, we dilate the ground truth mask gradually until it eventually expands into the whole image. Then, areas with the mask equal to one are filled with the corresponding PET and CT images. For input images without lesions, all channels of the output are zeros.

The advantage of this method is that to get the final segmentation result, the model needs to focus on a wider range of tissues first, and determine the location of the lesion step by step. This helps the model segment the lesion more accurately. The output channel expanded to the whole image can also be used to judge whether there are lesions in the input image, achieving the effect of classification. Another advantage of this method is that it resembles an autoencoder, which may replace the pre-training stage.

\subsection{Loss functions}
In terms of loss functions, we use Dicefocal Loss implemented by MONAI for the mask output channel and MSE Loss for other auxiliary channels. The weight of each loss is automatically adjusted using AutomaticWeightedLoss \cite{ref_article3}.

\subsection{Training and hyperparameter optimization}
Every training epoch, we go through all images in the whole data set, but images with lesions are sampled more frequently to prevent the model from always outputting zeros. The input image size is 32x160x160, so we use random crop to get the training images.

The network model was built with Pytorch and trained on an Nvidia TITAN RTX with 24 GB of memory. We use a batch size of 16. The optimizer is Adam. The initial learning rate is 3x10\textsuperscript{-4}. The learning rate is divided by 2.5 when the decrease of moving average training loss is less than 5x10\textsuperscript{-3}. When the learning loss is less than 10\textsuperscript{-5}, which usually takes 160 epochs, we reset the learning rate to 10\textsuperscript{-4} and use cosine annealing to train another 100 epochs.

We use LMDB dataset and compress pickle to speed up disk I/O. The size of the dataset is compressed from more than 300Gb to 52Gb.

\section{Result}
We evaluate our model using a validation set of 40 patients and the official preliminary test set, the result is shown in \textcolor{red}{Table. 1}

\begin{table}
\caption{Segmentation Result of our method.}\label{tab1}
\centering
\begin{tabular}{cccc} 
\hline
                                  & \textbf{~ ~DSC~ ~} & \textbf{~ ~False Positive~ ~} & \textbf{~ ~False Negative~ ~}  \\ 
\hline
\textbf{Validation set}           & 0.80  & 1.87 & 1.14    \\
\textbf{~ Preliminary~Test set~~} & 0.92               & 0.89                           & 0.53                            \\
\hline
\end{tabular}
\end{table}

\section{Conclusion}
The idea of step-by-step thinking is interesting. In this challenge, we use dilated masks as auxiliary channels to improve segmentation performance. But we still need more ablation experiments to prove the actual effect of this method.

\end{document}